\documentclass[preprint,aps,floats,showpacs]{revtex4}
\usepackage{graphicx}
\usepackage{epsfig}
\usepackage{pstricks}
\usepackage{pst-coil}
\usepackage{amsmath}

\def\be{\begin{equation}}
\def\ee{\end{equation}}
\def\bea{\begin{eqnarray}}
\def\eea{\end{eqnarray}}
\begin{document}
\title{\bf Bounds on neutrino masses from leptogenesis in 
type-II see-saw models}

\author{Narendra Sahu}
\email{narendra@phy.iitb.ac.in}
\author{S.~ Uma Sankar}
\email{uma@phy.iitb.ac.in}
\affiliation{Department of Physics, Indian Institute of 
Technology, Bombay, Mumbai 400076, India}                                                                       

\begin{abstract}
The presence of the triplet $\Delta_{L}$ in left-right 
symmetric theories leads to type-II see-saw mechanism 
for the neutrino masses. In these models, assuming a 
normal mass hierarchy for the heavy Majorana neutrinos, 
we derive a lower bound on the mass of the lightest of 
heavy Majorana neutrino from the leptogenesis constraint. 
From this bound we establish a consistent picture for the 
hierarchy of heavy Majorana neutrinos in a class of 
left right symmetric models in which we identify the 
neutrino Dirac mass matrix with that of Fritzsch type 
charged lepton mass matrix. It is shown that these 
values are compatible with the current neutrino 
oscillation data. 
\end{abstract}
\maketitle
%%%%%%%%%%%%%%%%%%%%%%%%%%%%%%%% 
\section{Introduction}
A plausible explanation of the observed baryon ($B$) asymmetry
of the Universe is that it arose from a lepton ($L$) asymmetry
~\cite{fukugita.86, luty.92, mohapatra.92, plumacher.96}. 
The conversion of the $L$-asymmetry to the $B$-asymmetry 
then occurs via the high temperature behavior of the $B+L$ 
anomaly of the Standard Model ($SM$)~\cite{krs.86}. This is an 
appealing route for several reasons. The extremely small 
neutrino masses suggested by the solar~\cite{solar} and 
atmospheric~\cite{atmos} neutrino anomalies and the 
KamLAND experiment~\cite{kamland}, point to the 
possibility of Majorana masses for the neutrinos generated by 
the see-saw mechanism~\cite{ge-ra-sl-ya}. This suggests the 
existence of new physics at a predictable high energy scale. 
Since the Majorana mass terms violate lepton number they can 
generate $L$-asymmetry. 

Early proposals along these lines relied on out-of-equilibrium
decay of the heavy Majorana neutrinos to generate the
$L$-asymmetry. In the simplest scenario a right-handed 
neutrino per generation is added to the $SM$~\cite{fukugita.86,
luty.92,plumacher.96}. They are coupled to left-handed 
neutrinos via Dirac mass matrix ($m_{D}$) which is assumed 
to be similar to charged lepton mass matrix~\cite{ge-ra-sl-ya}. 
Since the right handed neutrino is a singlet under $SM$ 
gauge group a Majorana mass term ($M_{R}$) for it can be 
added to the Lagrangian. Diagonalisation of neutrino mass 
matrix leads to two Majorana neutrino states per generation: 
a light neutrino state (mass $\sim m_{D}^{2}/M_{R}$) which is 
almost left handed and a heavy neutrino state (mass $\sim M_{R}$) 
which is almost right handed. This is called type-I see-saw 
mechanism in which the left handed fields do not have 
Majorana mass terms in the Lagrangian. 

It is desirable to consider neutrino masses in the context 
of grand unified theories ($GUTs$). The gauge groups of most 
of the GUTs contain the left-right symmetry group 
$SU(2)_L\times SU(2)_R$ as a subgroup~\cite{slansky_rep}. In 
such models Majorana masses, $M_L$, for left handed neutrinos 
occur in general, through the vacuum expectation value (VEV) 
of the triplet $\Delta_L$~\cite{magg-wet.80,wett.81,moh-senj.81,
laz-shf-wet.81,moha-susy-book.92}. In these models also there 
is a light and a heavy neutrino state per generation. The heavy 
neutrino state has mass $\sim M_{R}$ but the light neutrino mass 
is $\sim(M_{L}-m_{D}^{2}/M_{R})$. The presence of new 
scalars and their couplings, which give rise to $M_L$, can 
contain adequate $CP$-violation to accommodate $L$-asymmetry.
The two contributions to the light neutrino mass, $m_D^2/M_R$ 
and $M_L$ are called type-I and type-II terms respectively.
     
An additional grace of left-right symmetric models
~\cite{leftright_group} is that $B-L$ is a gauge symmetry 
in contrast to type-I models where $B-L$ conservation 
is ad-hoc. Because of $B-L$ is a gauge 
charge of this model, no primordial $B-L$ can exist. Further, 
the rapid violation of the $B+L$ conservation by the anomaly due 
to the high temperature sphaleron fields erases any $B+L$ 
generated earlier. Thus the lepton asymmetry must be produced 
entirely during or after the $B-L$ symmetry breaking phase 
transition.

Several authors~\cite{davidson&ibarra.02, buch-bari-plumcher-03, 
hamaguchi-etal-02, hamby-03} have recently dealt with the 
bound on the mass scale of lightest right handed neutrino $(M_1)$ in 
type-I see-saw models from the leptogenesis constraint. With the 
assumption of hierarchical mass spectrum of the heavy Majorana 
neutrinos, a common outcome was that $M_{1}\geq 10^{9}$ GeV. 
%An early attempt~\cite{antusch}, on the other hand, in type-II 
%see-saw models is made to derive a bound on the mass scale of 
%lightest right handed neutrino by assuming that both light and 
%heavy Majorana neutrino bases are diagonal simultaneously. It is 
%reported there that this bound can be reduced by an order of 
%magnitude in comparison to the type-I case. 
%%%%%%%%%%%%%%%%%%%%%%%%%%%%%%%%%%%%%%%%%%%%%%%%%%%%%%%%%%%%%%%%%%%
%%%%%%%%%%%%%%%%%%%%%% NOTE ON 6th NOVEMBER %%%%%%%%%%%%%%%%%%%%%%%
%%%%% REPLACE THE ABOVE SIX LINES BY THE FOLLOWING THREE LINES %%%
%%%%%%%%%%%%%%%%%%%%%%%%%%%%%%%%%%%%%%%%%%%%%%%%%%%%%%%%%%%%%%%%%%%
With the same assumption, a bound on $M_1$ was obtained in type-II
see-saw models in~\cite{antusch}, which is an order of magnitude
less than that in type-I case. 

In this paper we revisit the lower bound on the mass scale of 
lightest heavy Majorana neutrino~\cite{davidson&ibarra.02} 
due to its CP-violating decay to SM particles in generic 
type-II see-saw models. Recently it is claimed that the 
large atmospheric neutrino mixing can be achieved 
naturally in case of renormalisable $SO(10)$ theories 
if the type-II term dominates~\cite{baj-sen-vis.03,
bajcetal.04}. Therefore, it is interesting to extend 
our formalism to these models and derive an upper bound on 
the CP-asymmetry. In both cases it is shown that the mass 
scale of lightest right handed neutrino satisfies the 
constraint, $M_1\geq 2.5\times 10^{8}$GeV in order to 
produce the present baryon asymmetry of the Universe. 
On the other hand leptogenesis in models, where the type-II 
term was included in the neutrino mass matrix, was considered 
in the literature~\cite{joshipura_npb.01,joshipura_jhep.01}. 
However, the contribution to the CP-violating parameter, 
$\epsilon_1$, due to the triplet $\Delta_L$ in the loop, are 
not taken into account.  

Rest of our paper is organised as follows. In section II, we 
derive an upper bound on the $CP$-asymmetry in type-II see-saw 
models assuming that type-I and type-II terms are similar in 
magnitude. In section III, we discuss the upper bound on the 
mass scale of lightest right handed neutrino from the 
leptogenesis constraint. In light of current neutrino 
oscillation data in section IV a consistent picture for the 
heavy Majorana neutrino mass hierarchies are obtained in 
a class of left-right symmetric models in which we identify 
the neutrino Dirac mass matrix with that of charged lepton 
mass matrix~\cite{ge-ra-sl-ya}. Further we choose this 
matrix to be of Fritzsch type~\cite{fritzsch.79}. 
Finally in Section V, we put our summary and conclusions.
 
%%%%%%%%%%%%%%%%%%%%%%%%%%%%%%%%%%%%%%%%%%%
\section{Upper bound on $CP$-asymmetry in type-II models}
%%%%%%%%%%%%%%%%%%%%%%%%%%%%%%%%%%%%%%%%%%
Before proceeding with our analysis, we review 
the breaking scheme of $SO(10)$ grand unified theory 
through the left-right symmetric path~\cite{moh-senj.81, 
moha-susy-book.92}. This breaking can be accomplished by 
using a $\{126\}$ of $SO(10)$ as an intermediate. Under 
$SU(2)_{L}\otimes SU(2)_{R}\otimes SU(4)_{C}(= SU(2)_{L}\otimes 
SU(2)_{R}\otimes SU(3)_{C}\otimes U(1)_{B-L})$ its decomposition 
can be written as
\be
\{126\}=\Delta_{L}(3,1,10)+\Delta_{R}(1,3,10)+\Phi(2,2,15)
+\sigma(1,1,6),
\ee 
where $\sigma(1,1,6)$ is a singlet under $SU(2)_{L}\otimes 
SU(2)_{R}$ and has no role in neutrino mass generation. 
Since $126$ of $SO(10)$ contains a pair of triplets 
$\Delta_{L}$ and $\Delta_{R}$ and the bidoublet $\Phi$, 
left-right symmetry can be preserved at the intermediate 
level. As the right handed triplet $\Delta_{R}(1,3,10)$ 
gets a VEV $v_R$, left-right symmetry is broken to  
the SM symmetry. The scalars in both the triplets $\Delta_L$ 
and $\Delta_R$ acquire masses of the order $v_R$. At a lower 
scale, $\Phi$ gets a VEV $v$ breaking the SM symmetry to 
$U(1)_{em}$. This induces a small VEV $v_L$ for the neutral 
component of the triplet $\Delta_L$ \cite{moh-senj.81,
laz-shf-wet.81,hamb-senj.03}. The three VEVs are related by 
$v_L = \gamma v^2/v_R$, where $\gamma$ is a model dependent 
parameter which depends on the quartic couplings of the Higgs 
and can be as small as $10^{-4}$.

After the final symmetry breaking, the effective neutrino mass 
matrix is
\be
M_{\nu}=\bar{\nu}_{Li}\tilde{m}_{D_{ij}}\nu'_{Rj}+\frac{1}{2}
\left[ v_{L}\bar{\nu}_{Li}f_{ij}\nu^c_{Lj}+v_{R}\bar{\nu'}^{c}_{Ri}
f_{ij}\nu'_{Rj}\right]+H.C.
\label{Yukawa}
\ee
Because of $L \leftrightarrow R$ symmetry, the same 
symmetric Yukawa matrix $f_{ij}$ gives rise to Majorana masses 
for both left and right handed fields. The Dirac mass matrix is 
given by $\tilde{m}_{D}=\tilde{h}v$, where $\tilde{h}$ is the 
Yukawa matrix for neutrino Dirac masses.

The Majorana mass matrix for the right handed neutrinos can be 
diagonalized by making the following rotation on $\nu'_R$
\be
\nu_{R} = U_{R}^{\dagger}\nu_{R}^{'}. 
\ee
In this basis, we have
\bea
U_{R}^{T} f U_{R} & = & f_{dia}, 
\label{fdia} \\
h &=& \tilde{h}U_{R}.
\eea
In this rotated basis we get the mass matrix for the 
neutrinos  
\be
\begin{pmatrix}fv_{L} & m_{D}\\
m_{D}^{T} & M_{R}
\end{pmatrix},
\label{massmatrix}
\ee
where $M_{R}=f_{dia}v_{R}$ and $m_D = h v$. Diagonalising the 
mass matrix (\ref{massmatrix}) into $3\times3$ blocks
we get the mass matrix for the light neutrinos to be
\bea
m_{\nu} &=& fv_{L}-\frac{v^{2}}{v_{R}} h f_{dia}^{-1}h^T\nonumber\\
        &=&m_{\nu}^{II}+m_{\nu}^{I}.
\label{see-saw}
\eea 
Note that in contrast to the present case in type-I models, 
$m_{\nu}^{II}$ is absent. Diagonalization of the above light 
neutrino mass matrix $m_{\nu}$, through lepton flavour mixing 
PMNS matrix $U_{L}$, gives us three light Majorana neutrinos. 
Its eigenvalues are
\be
U_L^{\dagger} m_{\nu} U_L^*=dia(m_{1}, m_{2}, m_{3})\equiv D_m,
\label{diag}
\ee
where the masses are real.  
 
We assume a normal mass hierarchy for heavy Majorana neutrinos. 
In this scenario while the heavier neutrinos, $N_2$ and $N_3$, 
decay yet the lightest of heavy Majorana neutrinos is still 
in thermal equilibrium. Any asymmetry produced by the decay 
of $N_2$ and $N_3$ will be washed out by the lepton 
number violating interactions mediated by $N_1$. Therefore, the
final lepton asymmetry is given only by the CP-violating 
decay of $N_1$ to standard model leptons ($l$) and Higgs 
($\phi$). The CP-asymmetry in this scenario is given by 
\be
\epsilon_1=\epsilon_1^I+\epsilon_1^{II},
\label{epsilon}
\ee
where the contribution to $\epsilon_1^I$ comes from the 
interference of tree level, self-energy correction and the 
one loop radiative correction diagrams involving the heavier 
Majorana neutrinos $N_2$ and $N_3$. This contribution is the 
same as in type-I models~\cite{davidson&ibarra.02,
buch-bari-plumcher-03} and is given by 
\be
\epsilon_1^{I}=\frac{3M_1}{16\pi v^2}\frac{\sum_{i,j}Im
\left[ (h^{\dagger})_{1i}(m_{\nu}^I)_{ij}(h^*)_{j1}\right]}
{(h^{\dagger}h)_{11}}.
\label{epsilonI}
\ee
On the other hand the contribution to $\epsilon_1^{II}$ in 
equation (\ref{epsilon}) comes from the interference of tree 
level diagram and the one loop radiative correction 
diagram involving the triplet $\Delta_L$. It is given 
by~\cite{hamb-senj.03,antusch}
\be
\epsilon_1^{II}=\frac{3M_1}{16\pi v^2}\frac{\sum_{i,j}Im
\left[ (h^{\dagger})_{1i}(m_{\nu}^{II})_{ij}(h^*)_{j1}\right]}
{(h^{\dagger}h)_{11}}.
\label{epsilonII}
\ee
Substituting (\ref{epsilonI}) and (\ref{epsilonII}) in equation 
(\ref{epsilon}) we get the total CP-asymmetry 
\be
\epsilon_1 =\frac{3M_1}{16\pi v^2}\frac{Im(h^\dagger 
m_\nu h^*)_{11}}{(h^\dagger h)_{11}}.
\label{cpasym-1}
\ee
Using (\ref{diag}) in the above equation (\ref{cpasym-1}) 
we get 
\bea 
\epsilon_1 &=& \frac{3M_1}{16\pi v^2} \frac{ Im 
(h^\dagger U_L D_m U_L^T h^*)_{11}}{(h^\dagger h)_{11}}\nonumber\\
&=&\frac{3M_1}{16\pi v^2} \frac{\sum_i m_i Im(U_L^T h^*)_{i1}^2}
{\sum_i |(U_L^Th^*)_{i1}|^2}.
\label{cpasym-2}
\eea
With an assumption of normal mass hierarchy for the light 
Majorana neutrinos the maximum value of CP-asymmetry 
(\ref{cpasym-2}) can be given by
\be
\epsilon_{1,max} = \frac{3 M_1}{16\pi v^2}m_3.
\label{cpasym-3}
\ee 
Note that the above upper bound (\ref{cpasym-3}) for 
$\epsilon_1$ as a function of $M_1$ and $m_3$ was first 
obtained for the case of type-I see-saw 
models~\cite{davidson&ibarra.02}. However, the same 
relation holds in the case of type-II see-saw 
models also~\cite{antusch} {\it independent of the relative 
magnitudes of $m_{\nu}^{I}$ and $m_{\nu}^{II}$}. 
   
%%%%%%%%%%%%%%%%%%%%%%%%%%%%%%%%%%
\section{Estimation of Baryon Asymmetry}
%%%%%%%%%%%%%%%%%%%%%%%%%%%%%%%%%%%%%%%%%%
A net $B-L$ asymmetry is generated when left-right symmetry
breaks. A part of this $B-L$ asymmetry then gets converted to
$B$-asymmetry by the high temperature sphaleron transitions.
However these sphaleron fields conserve $B-L$. Therefore, the
produced $B-L$ will not be washed out, rather they will
keep on changing it to $B$-asymmetry. In a comoving volume
a net $B$-asymmetry is given by
\bea
Y_B &=& \frac{n_B}{s}=\frac{28}{79} \epsilon_1 Y_{N_1}\delta,
\label{B-asym}
\eea
where the factor $\frac{28}{79}$ in front~\cite{harvey.90} is the
fraction of $B-L$ asymmetry that gets converted to $B$-asymmetry.
Further  $Y_{N_1}$ is density of lightest right handed neutrino
in a comoving volume given by $Y_{N_1}=n_{N_1}/s$, where
$s=(2\pi^{2}/45)g_{*}T^{3}$ is the entropy density of the Universe
at any epoch of temperature $T$. Finally $\delta$ is the wash 
out factor at a temperature just below the mass scale of 
$N_1$. The value of $Y_{N_1}$ depends on the source of 
$N_1$. For example, the value of $Y_{N_1}$ estimated from 
topological defects~\cite{sahuetal.04} can be different from 
thermal scenario~\cite{plumacher.96, buch-bari-plumcher-03}. In
the present case we will restrict ourselves to thermal
scenario only.
 
Recent observations from WMAP show that the matter-antimatter
asymmetry in the present Universe measured in terms of
$(n_B/n_\gamma)$ is~\cite{spergel.03}                               
\be
\left(\frac{n_{B}}{n_{\gamma}}\right)_{0}\equiv
\left(6.1^{+0.3}_{-0.2}\right)\times 10^{-10},
\label{baryon-asym}
\ee
where the subscript $0$ presents the asymmetry today. 
Therefore, we recast equation (\ref{B-asym}) in terms of 
$(n_B/n_\gamma)$ and is given by  
\bea
\left(\frac{n_{B}}{n_{\gamma}}\right)_{0} &=& 7 (Y_{B})_{0}
\nonumber\\  
&=& 2.48 \epsilon_{1} Y_{N_1}\delta.
\label{baryon-asym1}
\eea
Substituting equation (\ref{cpasym-3}) in (\ref{baryon-asym1}) 
we get 
\be
\left(\frac{n_{B}}{n_{\gamma}}\right)_{0}\leq 2.48\left( 
\frac{3 M_1}{16\pi v^2} \right) m_3 Y_{N_1}\delta.
\label{baryon-asym2}
\ee 
The present neutrino oscillation data favours the bilarge 
neutrino mixing with the mass squared differences 
$\Delta m^2_{\rm atm}\equiv |m_3^2-m_1^2|\approx 
2.6\times10^{-3} eV^2$ and $\Delta m^2_{\rm sol}\equiv 
|m_2^2-m_1^2|\approx 7.1\times 10^{-5} eV^2$. Assuming 
a normal mass hierarchy ($m_3^2\gg m_2^2\gg m_1^2$) for 
the light Majorana neutrinos the above number gives $m_3\simeq
\left(\Delta m^2_{\rm atm}\right)^{1/2}\simeq 0.05 eV$. With 
this approximation we get from equation (\ref{baryon-asym2})
\be
M_1\geq 2.5\times 10^{8} GeV \left( \frac{10^{-2}}{Y_{N_1}\delta}
\right)\left(\frac{0.05eV}{m_3}\right).
\ee

For the above lower limit on the mass scale of lightest right 
handed neutrino, $M_1\geq 10^{8}$ GeV, we now proceed to find 
the plausible hierarchies $(M_2/M_1)$ and $(M_3/M_1)$ of the 
massive Majorana neutrinos in a model that are compatible with 
the current neutrino oscillation data. We check that the values 
we obtained are consistent with our assumptions.  
 
%%%%%%%%%%%%%%%%%%%%%%%%%%%%%%%%%%%%%%%%%%%%%%       
\section{Examining the consistency of f-matrix eigenvalues}
%%%%%%%%%%%%%%%%%%%%%%%%%%%%%%%%%%%%
The solar and atmospheric evidences of neutrino oscillations 
are nicely accommodated in the minimal framework of the 
three-neutrino mixing, in which the three neutrino flavours 
$\nu_{e}$, $\nu_{\mu}$, $\nu_{\tau}$ are unitary linear 
combinations of three neutrino mass eigenstates $\nu_{1}$, 
$\nu_{2}$, $\nu_{3}$ with masses $m_{1}$, $m_{2}$, $m_{3}$ 
respectively. The mixing among these three neutrinos determines 
the structure of the lepton mixing matrix \cite{mns-matrix} which
can be parameterized as
\be
U_{L}=\begin{pmatrix}
c_{1}c_{3} & s_{1}c_{3} & s_{3}e^{i\delta}\\
-s_{1}c_{2}-c_{1}s_{2}s_{3}e^{i\delta} & c_{1}c_{2}-s_{1}s_{2}s_{3}
e^{i\delta} & s_{2}c_{3}\\
s_{1}s_{2}-c_{1}c_{2}s_{3} & -c_{1}s_{2}-s_{1}c_{2}s_{3}e^{i\delta} &
c_{2}c_{3}\end{pmatrix} dia(1, e^{i\alpha}, e^{i(\beta +\delta)}),
\label{mns-matrix}
\ee
where $c_{j}$ and $s_{j}$ stands for $cos \theta_{j}$
and $sin \theta_{j}$. The two physical phases
$\alpha$ and $\beta$ associated with the Majorana
character of neutrinos are not relevant for
neutrino oscillations \cite{bilenkyetal.80} and will be
set to zero here onwards. While the Majorana phases
can be investigated in neutrinoless double beta decay
experiments \cite{rodejohan_npb.01}, the CKM-phase
$\delta \in [-\pi, \pi]$ can be investigated in
long base line neutrino oscillation experiments. For 
simplicity we set it to zero, since we are
interested only in the magnitudes of elements of $U_{L}$.
The best fit values of the neutrino masses and mixings from
a global three neutrino flavors oscillation analysis are 
\cite{gonzalez-garcia_prd.03} 
\be
\theta_{1}=\theta_{\odot}\simeq 34^\circ, ~~\theta_{2}=\theta_{atm}
=45^\circ, ~~\theta_3 \leq 13^\circ,
\label{bestfit-theta}
\ee
and
\bea
\Delta m_{\odot}^{2}= m_2^2 - m_1^2 & \simeq & m_2^2 =
7.1\times 10^{-5} \ {\rm eV}^{2}\nonumber\\
\Delta m_{atm}^{2}= m_3^2 - m_1^2 & \simeq & m_3^2 =
2.6\times 10^{-3} \ {\rm eV}^{2}.
\eea
From equation (\ref{see-saw}) and (\ref{diag}) we have 
\be
f=\frac{1}{v_L}\left[(U_L D_m U_L^T)+\frac{v^{2}}{v_{R}} 
(h f_{dia}^{-1}h^T)\right].
\label{f-matrix}
\ee
We now assume a hierarchical texture for the Majorana Yukawa 
coupling to be 
\be
f_{dia}= \frac{M_{1}}{v_{R}}\begin{pmatrix}1 & 0 & 0\\
0 & \alpha_{A} & 0\\
0 & 0 & \alpha_{B} \end{pmatrix},
\label{fdia-texture}
\ee  
where $1 \ll \alpha_{A}=(M_{2}/M_{1}) < \alpha_{B}=
(M_{3}/M_{1})$. We identify the neutrino Dirac Yukawa
coupling $h$ with that of charged leptons~\cite{ge-ra-sl-ya}. 
Further we assume it to be of Fritzsch type~\cite{fritzsch.79}
\be
h=\frac{(m_{\tau}/v)}{1.054618}
\begin{pmatrix}
0 & a & 0\\
a & 0 & b\\
0 & b & c\end{pmatrix}.
\label{h-texture}
\ee
By choosing the values of a,b and c suitably one can get
the hierarchy for charged leptons. In particular, the values
\be
a=0.004,~~~ b=0.24 ~~~{\rm and}~~~ c=1
\label{abcvalues}
\ee
give the mass hierarchy of charged leptons. For this set 
of values the mass matrix h is normalized with respect to the
$\tau$-lepton mass. The values of $a, b$ and $c$
are rougly in geometric progression. They can be expressed
in terms of the electro-weak gauge coupling $\alpha_{w}=
\frac{g^{2}}{4\pi}=\frac{\alpha}{sin^{2}\theta_{w}}$. In
particular $a=2.9 \alpha_{w}^{2}$, $b=6.5 \alpha_{w}$
and $c=1$. Here onwards we will use this set of values
for the parameters of $h$. 

Substituting (\ref{fdia-texture}) and (\ref{h-texture}) in 
equation (\ref{f-matrix}) we get 
\be
f=(\frac{eV}{v_{L}})\left[((U_L D_m U_L^T/eV)+\frac{4}
{(1.054165)^{2}}\frac{1}{(M_{1}/{\rm GeV})}
\begin{pmatrix}
\frac{a^{2}}{\alpha_{A}} & 0 & \frac{ab}{\alpha_{A}}\\
0 & a^{2}+\frac{b^{2}}{\alpha_{B}} & \frac{bc}{\alpha_{B}}\\
\frac{ab}{\alpha_{A}} & \frac{bc}{\alpha_{B}} & \frac{b^{2}}
{\alpha_{A}}+\frac{c^{2}}{\alpha_{B}}
\end{pmatrix} \right],
\label{maj-yukawa}
\ee
For the demonstration purpose we choose a typical value of 
$M_1=2.5\times 10^{8}$ GeV. Now by suitably choosing the 
parameters $m_{1}=1.0 \times 10^{-4}eV$, $\alpha_{A}=51$, 
$\alpha_{B}=191$, $\theta_{13}=10^\circ$ we get
\be
f_{dia}=\frac{6.42 \times 10^{-4}eV}{v_{L}}
\begin{pmatrix}
1 & 0 & 0\\
0 & 50.98 & 0\\
0 & 0 & 190.47 \end{pmatrix}.
\label{fdia-cal}                          
\ee
Thus, for the above values of $m_1$ and $M_1$, the
assumed hierarchies of right-handed neutrino masses
are consistent with global low energy neutrino data. 
Further we emhasize that the value of $m_1$ for which 
the consistency is obtained is compatible with our 
assumption that 
\be 
m_1^2\ll m_3^2\equiv \Delta m^2_{atm}.
\ee

Comparing equation (\ref{fdia-cal}) with (\ref{fdia-texture})
one can get
\be
\frac{M_{1}}{v_{R}}=\frac{6.42 \times 10^{-4}eV}{v_{L}}.
\ee
This implies that $v_{R}=O(10^{11})$ GeV for $v_{L}=0.1$ eV. 
These values of $v_{L}$ and $v_{R}$ are compatible with genuine
see-saw $v_{L}v_{R}=\gamma v^{2}$ for a small value of
$\gamma\simeq O(10^{-3})$. Although in the literature 
frequently it is stated $\gamma$ is to be order of unity, we 
see, however, here that it is required to be $O(10^{-3})$ for the 
consistency. See, for example, the recently proposed type-II 
see-saw models in which $\gamma=v_L v_R/v^2 \geq O(10^{-4})$
~\cite{nimai.04}.

Here we demonstrated the consistency of our choice of the
matrix $f$ with the current neutrino data for the minimum 
value of $M_1$. For higher values of $M_1$, to be consistent 
with $1<<\alpha_A <\alpha_B$, one can choose appropriate values 
of $m_1\leq 10^{-3}$ eV and $\theta_{13}\leq 13^{\circ}$ in 
equation (\ref{maj-yukawa}) which will reproduce the correct eigenvalues of 
the matrix $f$. 

%%%%%%%%%%%%%%%%%%%%%%%%%%%%%%%%%%%%%%%%%
\section{Conclusion}
In this work we proved the universality of the upper bound on the 
$CP$-violating parameter $\epsilon_{1}$ in general 
type-II see-saw models. Irrespective of any assumption 
regarding the magnitude of type-I and type-II terms we found 
that same bound holds for all cases. Assuming a normal 
mass hierarchy for the massive Majorana neutrinos we 
demonstrated that in a thermal scenario the present baryon 
asymetry can be explained for all values of $M_1\geq 
O(10^8)GeV$ in case of left-right symetric models. 
We also consider a class of left-right symmetric models 
in which we assume the neutrino Dirac masses are of the 
Fritzsch type. In such models we found that plausible 
hierarchies for the massive Majorana neutrinos can be obtained 
which are compatible with the current neutrino oscillation data.

%%%%%%%%%%%%%%%%%%%%%%%%%%%%%%%%%%%%%%%%%
\section*{Acknowledgment}It is our pleasure to thank Prof. U. A.
Yajnik for his critical comments on the draft. We also thank 
Prof. R. N. Mohapatra and Prof. M. K. Parida for their interest 
and valuable discussions during the WHEPP-8 held at IIT Bombay. 
%%%%%%%%%%%%%%%%%%%%%%%%%%%%%%%%%%%%%%%%%%%

\end{document}